\documentclass[12pt, draftclsnofoot, onecolumn]{IEEEtran}
\usepackage{amsmath,epsfig,algorithm,algorithmic}
\usepackage{multirow,caption,subfigure}

\newtheorem{thm}{Theorem}[section]
\newtheorem{lem}{Lemma}[section]

\ifCLASSINFOpdf
\else
\fi
\begin{document}
%
\title{QoE-based  MAC Layer Optimization for Video Teleconferencing over WiFi}
%
%
%

\author{Tianyi~Xu,~\IEEEmembership{Member,~IEEE,}
        Liangping~Ma,~\IEEEmembership{Senior~Member,~IEEE,}
        and~Gregory~Sternberg,~\IEEEmembership{Senior~Member,~IEEE}
\thanks{T. Xu and L. Ma are with InterDigital Communications, Inc., San Diego, CA 92121, USA (e-mail: \{tianyi.xu, liangping.ma\}@interdigital.com).}
\thanks{G. Sternberg is with InterDigital Communications, Inc., King of Prussia, PA 19406, USA (e-mail: gregory.sternberg@interdigital.com).}
}

\maketitle

\begin{abstract}
In IEEE 802.11, the retry limit is set the same value for all packets. In this paper, we dynamically classify video teleconferencing packets based on the type of the video frame that a packet carries and the packet loss events that have happened in the network, and assign them different retry limits. We consider the IPPP video encoding structure with instantaneous decoder refresh (IDR) frame insertion based on packet loss feedback. The loss of a single frame causes error propagation for a period of time equal to the packet loss feedback delay. To optimize the video quality, we propose a method to concentrate the packet losses to small segments of the entire video sequence, and study the performance by an analytic model. Our proposed method is implemented only on the stations interested in enhanced video quality, and is compatible with unmodified IEEE 802.11 stations and access points in terms of performance. Simulation results show that the performance gain can be significant compared to the IEEE 802.11 standard without negatively affecting cross traffic.
\end{abstract}

\begin{IEEEkeywords}
QoE, packet priority, video teleconferencing, IEEE 802.11, retry limit
\end{IEEEkeywords}

%
\IEEEpeerreviewmaketitle

\section{Introduction}\label{sec:intro}
%
%
%
%

\IEEEPARstart{I}{n} current real-time video applications, such as video teleconferencing, the IPPP video encoding structure is widely used to satisfy stringent delay constraints. The first frame of the video sequence is intra-coded, and each of the other frames is encoded by using the immediately preceding frame as the reference for motion compensated prediction. When transmitted in a lossy channel, a packet loss affects not only the corresponding frame but also the subsequent frames, which is called error propagation. To deal with packet losses, macroblock (MB) intra refresh \cite{Wiegand03wireless} may be used, where some MBs of each frames are intra-coded. This can alleviate the error propagation at the expense of lower coding efficiency.

In this paper, we consider that the video destination feeds back the packet loss information to the video encoder to trigger the insertion of an instantaneous decoding refresh (IDR) frame \cite{Wiegand03wireless}, which is intra-coded, so that the subsequent frames are free of error propagation. This is one of the reactive packet loss mitigation methods used in WebRTC \cite{webrtc}. Specifically, the packet loss information can be sent by the receiver via an RTP Control Protocol (RTCP) \cite{RTCP} packet containing the index of the frame to which the lost packet belongs. After receiving this information, the video encoder decides whether the packet loss creates a new error propagation interval. If the index of the frame to which the lost packet belongs is less than the index of the last IDR frame, the video encoder will do nothing. In this case, the packet loss occurs during an existing error propagation interval, and a new IDR frame has already been generated which will stop the error propagation. Otherwise, the packet loss creates a new error propagation interval, and the video encoder encodes the current frame in the intra mode to stop the error propagation.
The duration of error propagation depends on the feedback delay, which is at least a round trip time (RTT) between the video encoder and decoder. Another approach to alleviating error propagation is recurring IDR frame insertion \cite{Rapaport13}, where a frame is intra-coded after every fixed number of P frames, which is not considered in this paper.

In IEEE 802.11 MAC, when a transmission is not successful, a retransmission will be performed until the retry limit is exceeded. The retry limit is the maximum number of transmission attempts for a packet \cite[pp. 2134]{Wlan80211retry}, and a packet that could not be transmitted after this many attempts is discarded by the MAC. The IEEE 802.11 standard defines two retry limits: short retry limit for the packets with a length less than or equal to the RTS/CTS threshold, and long retry limit for the packets with a length greater than the RTS/CTS threshold \cite{Wlan80211retry}. In this paper, the use of RTS/CTS is disabled as is often seen in practice, and we only consider the short retry limit, which is denoted by $R$.

The importance of a video packet depends on not only the type of frames that a video packet carries but also the events that have happened in the network. Network events include packet losses or excessive delays. For example, for a given P frame, being the second lost packet may not have as much impact as being the first lost packet. The dynamic nature of network events makes the importance of a video packet also dynamic. The impact of network events on the video quality is often overlooked in most of the existing resource allocation algorithms. As a result, packet differentiation is static and depends only on the video encoding structure, such as recurring IDR frame insertion and scalable video coding (SVC) \cite{Wiegand07SVC}. For instance, SVC separates video packets into different substreams based on which layer a video packet belongs to, and conveys the priority level information of the substreams to the network, which then allocates more resources to the substreams with higher priorities. The prioritization based on SVC is static, as the priority of a video packet is determined at the time of video encoding and fixed throughout the lifetime of the video packet.

We propose a MAC layer optimization method that adapts the retry limit to both video frame types and the network events. According to the importance of the video packets, we dynamically assign retry limits; less important video packets are assigned a lower retry limit, and the saved retransmission opportunities or the earned credit is shifted to the more important video packets without increasing the total contention to the competing traffic. This idea is reminiscent of the one in \cite{Liangping11}, where the secondary spectrum users earn credit by assisting the transmission of the primary spectrum users and consume the credit in accessing the spectrum at a later time.

We also present an analytic model to evaluate the performance of our proposed method. In the literature, model-based throughput analyses for the IEEE 802.11 standard are proposed in \cite{Bianchi00}-\cite{Chat03}. In this paper, our focus is on the impact on the video quality resulting from our proposed MAC layer optimization method. Considering the transmission of cross traffic, a compatibility condition is also required to guarantee that cross traffic will not be negatively affected. By using simulations, we show that the throughput of cross traffic remains almost the same compared to that for the scenario where MAC layer optimization is not used. The proposed method and the analytic model are investigated under the assumption that no forward error correction (FEC) is performed. However, as we show, it is not difficult to modify the proposed method to be compatible with FEC. 

The remainder of this paper is organized as follows. Section \ref{sec:motivation} gives the motivation of our proposed method. Section \ref{sec:prop_method} describes our proposed method. Section \ref{sec:model} presents an analytic model. Section \ref{sec:simu} gives the simulation results. Section \ref{sec:conl} concludes this paper.

\section{Motivation and QoE metric}\label{sec:motivation}

\begin{figure}[t]
    \centering
    \includegraphics[width=0.7\textwidth]{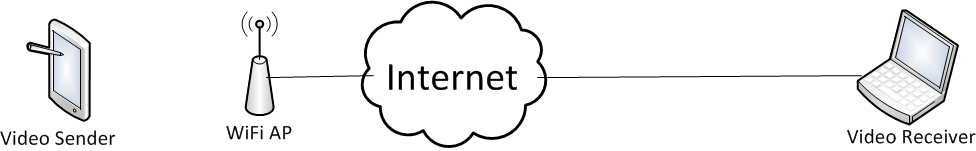}
    \caption{Video teleconferencing over WiFi}
    \label{fig:topo_s}
\end{figure}

In video teleconferencing, one of the most important communication networks is that the video sender connects through WiFi to the core Internet, and eventually connects to the video receiver, as illustrated in Fig. \ref{fig:topo_s}. Considering that packet losses occur most likely in the WiFi link, optimization in the WiFi link may improve the video receiver's quality of experience (QoE) significantly. Modification at the WiFi access point (AP) is difficult to implement, and often affects other users' performance in the same WiFi network. Thus, in this paper, our objective is to design a MAC layer optimization at the video sender to improve the QoE at the video receiver.

There are QoE models in the literature \cite{Khan}-\cite{Venkat09}. These models are essential to QoE-driven network resource allocation \cite{Jiang12}-\cite{Feng12}. However, most of the QoE models assume the availability of QoS parameters such as the video bit rate, frame rate and packet loss probability. These parameters are not suitable for our MAC layer optimization, because the video bit rate and frame rate are controlled by the video teleconferencing applications, and the packet loss probability is determined by the wireless channel condition and the cross traffic from other users in the same WiFi network.

\begin{figure}[t]
    \centering
    \includegraphics[width=0.6\textwidth]{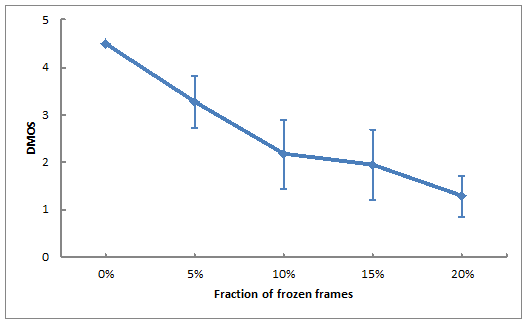}
    \caption{average DMOS of video sequences with different fractions of frozen frames and the $95\%$ confidence intervals}
    \label{fig:dmos}
\end{figure}

Instead, our proposed method tries to reduce the number of frozen frames to achieve better QoE. To avoid visual artifacts, some video teleconferencing applications, such as FaceTime \cite{Facetime}, Hangouts \cite{Hangouts} and Skype \cite{Skype}, choose to present the most recent error-free frame instead of erroneous frames. The video receiver freezes the video during an error propagation period, which is called a frozen interval, and the frames presented during this period are called frozen frames.
In this paper, we consider the freeze time as the performance metric. Given a constant frame rate, calculating the freeze time is equivalent to counting the number of frozen frames. It is desirable to establish the relationship between the fraction of frozen frames to the mean opinion score (MOS), where the frozen frames occur in groups, each lasing for a time period equal to the packet loss feedback delay. Therefore, we perform a subjective experiment to characterize how the fraction of frozen frames affects the QoE. We adopt the single stimulus absolute category rating with hidden reference (ACR-HR) method to obtain the quality scores \cite{wolf}. Ten observers are first asked to watch a training session, containing two Basketball Passing video sequences with no frozen frames and the most frozen frames, respectively. Then, with the training session in mind, the observers view five video sequences in a random order, all of which are the Foreman sequences but with the fraction of frozen frames varying from $0\%$ to approximately $20\%$. These video sequences are generated from the packet losses obtained in the OPNET simulations in Section \ref{sec:simu}. At the end of each video sequence, the observers are asked to rate the video according to the ITU five-point quality scale, where the scores 1 to 5 stand for the following quality levels: very annoying, annoying, slightly annoying, perceptible but not annoying and imperceptible, respectively. With hidden reference removal \cite{wolf}, the difference mean opinion score (DMOS) is obtained by subtracting the observer's rating of the reference video (i.e., the video with no frozen frames) from the observer's rating of other videos, and 5 is added to the DMOS to make it nonnegative. In Fig. \ref{fig:dmos}, we show the average DMOS of video sequences with different fractions of frozen frames as well as the $95\%$ confidence intervals. Clearly, by decreasing the fraction of frozen frames, we improve the viewer's subjective experience, especially when the fraction of frozen frames is less than $10\%$, which is the motivation of our proposed method. Similar subjective test results are also presented in \cite{Globalsip,Mohamed02}. In \cite{Globalsip}, it is claimed that videos with higher rebuffering frequency or longer rebuffering duration have worse QoEs compared to those with fewer rebufferings or shorter durations. In \cite{Mohamed02}, it shows that the QoE decreases as the packet lost rate increases, where the packet loss is the cause of frozen frames.

\section{QoE-based Optimization}\label{sec:prop_method}

\begin{figure}[t]
    \centering
    \includegraphics[width=0.6\textwidth]{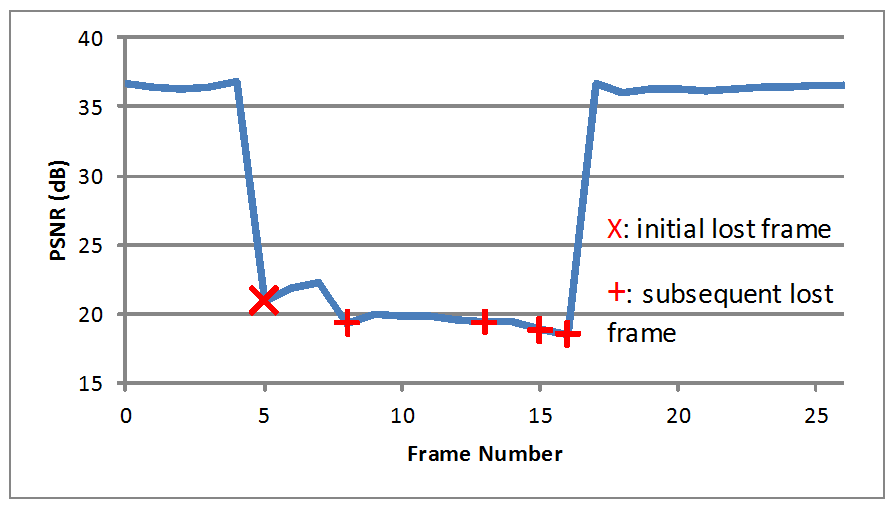}
    \caption{PSNR drop caused by packet losses}
    \label{fig:psnr}
\end{figure}

In the IEEE 802.11 standard, the retry limits are the same for all packets. This is not optimal for video teleconferencing traffic. To illustrate, we use the Foreman video sequence to show a special property of the IPPP encoded video sequences in the presence of packet losses in Fig. \ref{fig:psnr}. The video quality is measured by PSNR, the widely used objective video quality metric. Note that the use of PSNR here is only for the purpose of illustration. We understand that generally PSNR does not correlate well with QoE. However, a drop of as much as 10dB in PSNR does mean a significant drop in QoE. Later in the paper, we will evaluate the video quality by the video freeze time, which highly correlates with QoE as we have shown in Fig. \ref{fig:dmos}. The loss of frame 5 causes all the subsequent decoded P frames to be erroneous until the next IDR frame, and the video quality stays low regardless of whether the subsequent frames are received successfully. Thus, the transmission of these frames is less important to the video quality, and we may consider lowering the retry limit for them. In our proposed method, we classify the video frames into three priority categories, and assign retry limit $R_i$ for the video frames with priority $i$ $(i= 1,2,3)$, where priority 1 is the highest priority and $R_1>R_2=R>R_3$. First, an IDR frame and subsequent frames are assigned retry limit $R_1$, until a frame is lost or the compatibility constraint (to be discussed in detail later) is violated. After generating an IDR frame, we want the decoded video sequence at the receiver to be error-free as long as possible. Otherwise, if the network drops a frame shortly after the IDR frame, the video quality will decrease dramatically and remain poor until a new IDR frame is generated, which will take at least 1 RTT. The benefit of an IDR frame that is quickly followed by a packet loss will then be limited to a few video frames. Therefore, we prioritize not only the IDR frame but also the frames subsequent to the IDR frame. When the MAC layer discards a packet because it has reached the retry limit, the subsequent frames are assigned the smallest retry limit $R_3$ until a new IDR frame is generated, because a higher retry limit would not improve the video quality anyway. All the other frames  are assigned retry limit $R_2$, which is the same as the one in the original IEEE 802.11 standard. The reason behind this setting is to simplify the analytic model and the implementation. Indeed, it is interesting to find out in the analytic model that very few frames are assigned with the retry limit $R_2$.

Notice that the proposed method can be also implemented with FEC to improve video quality. However, with FEC, when a packet is discarded by the MAC layer, the receiver may still be able to decode the video frame because of the redundant FEC packets, and thus, the MAC layer does not have to lower the retry limit of the subsequent frames. Instead, the MAC layer only reduces the retry limit of the subsequent frames when the video frame cannot be successfully decoded by the receiver. In order to decide whether a video frame is decodable, the MAC layer has to know how the FEC packets are generated, such as the beginning of each FEC code block and FEC rate \cite{FEC}. This information can be placed in the extended RTP header as outlined in \cite{RTP}. The MAC layer then reads the information by deep packet inspection. In case of encryption, this information is still transmitted in plaintext in the optional SRTP extension field \cite{SRTP}. The proposed method can be easily generalized to the case when FEC is performed, and the analytic model is similar. For convenience, in the following sections, we only discuss the proposed method without FEC.

To facilitate the adoption of our proposed method, we impose a compatibility constraint on our design. In order to make sure that the performance of other access categories (ACs) is not negatively affected by optimizing the retry limits for the video packets, we try to maintain the same total number of transmission attempts of a video sequence before and after our proposed method is applied. In our study, instead of monitoring the actual number of transmission attempts, we estimate the average number of transmission attempts for the video packets as follows.

Let $p$ be the collision probability of a single transmission attempt at the MAC layer of the video sender when our proposed method is used. We adopt the assumption in \cite{Bianchi00}-\cite{Chat03} that $p$ is a constant and independent for every packet, regardless of the number of retransmissions.
In our proposed optimization method, the probability $p$ is monitored at the MAC layer, and then used as an approximation of collision probability when only the unmodified IEEE 802.11 standard is used.
The probability that a transmission still fails after $r$ attempts is $p^r$. For a packet with retry limit $R$, the average number of transmission attempts is given by
\begin{eqnarray}\label{eqn:avetrans}
\sum^R_{r=1}r\cdot p^{r-1}(1-p)+R\cdot p^{R}=\frac{1-p^R}{1-p},
\end{eqnarray}
where $p^{r-1}(1-p)$ is the probability that a packet is successfully transmitted after $r$ attempts, and $p^R$ in the second term on the left hand side is the probability that the transmission still fails after $R$ attempts.
For convenience, let $p_i=p^{R_i}$ for $i=1,2,3$, where $p_i$ is the packet loss rate when the retry limit is $R_i$. Since $R_1>R_2>R_3$, we have that $p_1<p_2<p_3$.


Let $M$ be the total number of video packets received by the MAC layer up to now for the case where all stations and the access point (AP) use the unmodified IEEE 802.11 standard, and then the average number of transmission attempts for the video sequence is given as
\begin{eqnarray*}
\frac{1-p_0}{1-\tilde{p}} M
\end{eqnarray*}where $\tilde{p}$ is the collision probability when the unmodified IEEE 802.11 standard is used at the video sender and $p_0=\tilde{p}^R$. 
Let $M_i$ $(i=1,2,3)$ be the total number of video packets received by the MAC layer with retry limit $R_i$ when our proposed method is used at the video user. Given the maximum transmission unit (e.g., the maximum MSDU size for WiFi is 2304 bytes \cite{Wlan80211retry}), the video packets corresponding to a video frame received at the MAC layer have the same sizes except the last one. Then it is reasonable to assume that for every video packets, every single transmission attempt has the same channel occupancy time. To guarantee that our proposed method does not increase the total channel occupancy, we require that the total number of transmission attempts does not increase after we adjust packet retry limits, i.e.,
\begin{eqnarray}\label{eqn:compa_ineqn_ori}
\frac{1-p_0}{1-\tilde{p}} M\geq\sum^3_{i=1}\frac{1-p_i}{1-p}M_i,
\end{eqnarray}
However, when our proposed method is used, both $\tilde{p}$ and $M$ are unavailable. Instead, we consider the following constraint
\begin{eqnarray}\label{eqn:compa_ineqn_alg}
\frac{1-p^R}{1-p}(M_1+M_2+M_3)\geq\sum^3_{i=1}\frac{1-p_i}{1-p}M_i.
\end{eqnarray}

\begin{figure}[t]
    \centering
    \includegraphics[width=0.65\textwidth]{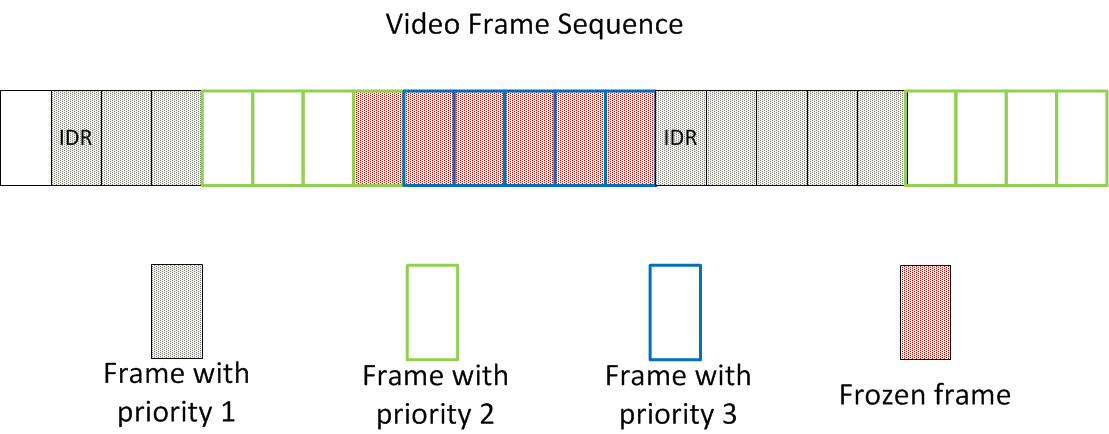}
    \caption{Video frame priorities assigned by our proposed method}
    \label{fig:frame_priority}
\end{figure}

Our proposed method is implemented at the MAC layer of the video sender with a linear computational complexity, and can be summarized as follows (see Fig. \ref{fig:frame_priority}): The IDR frames are always assigned priority 1. For a subsequent frame, if its preceding frames are transmitted successfully, it will be assigned priority 1 as long as the compatibility requirement (\ref{eqn:compa_ineqn_alg}) is satisfied. When the compatibility requirement is violated, the MAC assigns priority 2 to the current frame and the subsequent frames until a packet is dropped because of exceeding the retry limit. When a packet with priority 1 or 2 is dropped, the subsequent frames will be assigned priority 3 until the next IDR frame. The number of consecutive frames with priority 3 is determined by the duration of error propagation, which is at least 1 RTT. The details of the algorithm are shown in Algorithm \ref{alg1}.
The accumulative number of packets $M_i$ are calculated from the beginning of the video sequence. However, when the video duration is large, we may count the accumulative number of packets during a certain time period.
 
\begin{algorithm}[h]
\caption{Assign the priority to the video frames, and update the numbers of packets $M_i$}\label{alg1}
\begin{algorithmic}[1]
\STATE {Initialize $M_i=0$; the priorities of the current frame and the last frame, $q, q_0=0$, respectively};
\STATE {When a packet arrives from the higher layer, }\label{pckt_come}
\STATE {\textbf{IF} it belongs to a new video frame, then}
\STATE \quad{\textbf{IF} it belongs an IDR frame, then let $q=1$};
\STATE \quad{\textbf{ELSE IF} $q_0==3$, then let $q=3$};
\STATE \quad{\textbf{ELSE IF} at least one packet of last frame is dropped, then $q=3$};
\STATE \quad{\textbf{ELSE IF} $q_0==2$, then $q=2$};\label{step6}
\STATE \quad{\textbf{ELSE IF} inequality (\ref{eqn:compa_ineqn_alg}) is satisfied, then $q=1$};
\STATE \quad{\textbf{ELSE} $q=2$};\label{step8}
\STATE \quad{\textbf{END IF}}
\STATE \quad{$q_0=q$};
\STATE {\textbf{END IF}}
\STATE {$M_q=M_q+1$};\label{lin:update_M}
\STATE {Transmit the packet according to retry limit $R_q$, and update the collision probability $p$};\label{lin:12}
\STATE {Back to \ref{pckt_come} until the video session ends.}
\end{algorithmic}
\end{algorithm}

Notice that according to Steps \ref{step6}-\ref{step8} in Algorithm \ref{alg1}, a frame is assigned priority 2 only when the previous frame was assigned priority 2 or inequality (\ref{eqn:compa_ineqn_alg}) is violated. Hence, when inequality (\ref{eqn:compa_ineqn_alg}) is always satisfied, no frame will be assigned priority 2. In Section \ref{sec:model}, we will use this property to show that after the beginning of the video sequence, all frames are assigned priority 1 or 3.

Although increasing the retry limit for the packets with high priority may also increase the video latency, it only increases the actual number of transmission attempts for very few packets, which failed in the previous $R$ transmission attempts. Moreover, as we show in Lemma \ref{lem:pckt}, our proposed method reduces the total packets transmitted, and eventually reduces the packet delay, which is confirmed later by the simulation results presented in Section \ref{sec:simu}.

\section{Analytic Model}\label{sec:model}

Assume that for every IDR  and non-IDR video frame, $d$ and $d'$ packets with the same size are transmitted by the MAC layer of the video sender, respectively, where $d>d'$. Let $N$ be the total number of frames encoded thus far, and $n$ be the expected number of packets when the unmodified IEEE 802.11 standard is used at the video sender. For fair comparison, we consider the same number $N$ of video frames for the case where our proposed method is used.  Using our proposed method, we assign one of the three priorities to a frame, and denote by $n_i$ the expected number of packets with priority $i$. Notice that $n$ and $n_1+n_2+n_3$ are not necessarily equal, because the numbers of IDR frames in these two scenarios may be different. In the following analysis, we are only interested in the case where $N$ is large enough and consider the expected number of packets. 
The inequality (\ref{eqn:compa_ineqn_ori}) and (\ref{eqn:compa_ineqn_alg}) can be approximated by the following inequalities:
\begin{eqnarray}
\frac{1-p_0}{1-\tilde{p}} n\geq\sum^3_{i=1}\frac{1-p_i}{1-p}n_i\label{eqn:compat}\\
\frac{1-p^R}{1-p}(n_1+n_2+n_3)\geq\sum^3_{i=1}\frac{1-p_i}{1-p}n_i\label{eqn:compat_alg}
\end{eqnarray}

Considering a constant frame rate, we suppose that $D$ is the number of frames sent during a feedback delay. When a packet is lost in transmission, the packet loss information is received at the video source a feedback delay after the packet was sent. Then a new IDR frame is generated immediately, which is the $D$-th frame after the frame to which the lost packet belongs. $D-1$ frames are affected by error propagation. Even when the feedback delay is short, at least the frame to which the lost packet belongs is erroneous, so we assume that $D\geq1$, and call the interval containing the $D$ frozen frames the frozen interval.

We assume that the packet loss probability $p_0$ is so small that in each frozen interval, there is only one packet loss (the very first one), when the unmodified IEEE 802.11 standard is used at the video user. Then, the number of independent frozen intervals is equal to the number of lost packets, which is $p_0n$ in an $n$-packet video sequence. Thus, the total expected number of erroneous frames, i.e., frozen frames, is given by
\begin{equation}
N_f=p_0nD.\label{eqn:errfrm_ori}
\end{equation}

Using our proposed method, every frozen interval begins with an erroneous frame with priority 1 or 2, which is followed by $D-1$ frames with priority 3. The numbers of lost packets with priority 1 and 2 are $p_1n_1$ and $p_2n_2$, respectively. So, when our proposed method is used, the total expected number of frozen frames is
\begin{equation}\label{eqn:errfrm_prop}
N'_f= (p_1n_1+p_2n_2)D.
\end{equation}

Notice that the frames with priority 3 only appear in frozen intervals, and each is encoded into $d'$ packets. Each frozen interval contains $D$ frames, $D-1$ of which are assigned priority 3. Thus, the total expected number of packets with priority 3 is given by
\begin{eqnarray}\label{eqn:pkt3}
n_3= \frac{D-1}{D}N'_fd'.
\end{eqnarray}
When $D=1$ (corresponding to very short feedback delays), only one frame (the frame to which the lost packet belongs) is transmitted in the frozen interval, and the next frame is an IDR frame which stops the frozen interval. In this case, no frame is assigned priority 3, and $n_3=0$.

Denote by $n'_I$ the number of packets which belong to IDR frames, when our proposed method is used. Notice that except for the first IDR frame, other IDR frames appear only after the ends of frozen intervals, and each is encoded into $d$ packets. Thus, the total number of packets belonging to IDR frames is given by
\begin{eqnarray}\label{eqn:pkt_IDR}
n'_I=(\frac{N'_f}{D}+1)d
\end{eqnarray}

Before presenting the main theorem, we first introduce the following lemmas. 
\begin{lem}\label{lem:pckt}
If $\tilde{p}\geq p$, when our proposed method is used at the video sender, the expected number of packets is less than that when the original IEEE 802.11 standard is used at the video sender, i.e.,
\begin{eqnarray}
n_1+n_2+n_3<n.\label{eqn:pckt}
\end{eqnarray}
\end{lem}

\begin{lem}\label{lem:comp1}
When $\tilde{p}\geq p$, we have
\begin{eqnarray*}
\frac{1-p_0}{1-\tilde{p}} n\geq\sum^3_{i=1}\frac{1-p_i}{1-p}n_i
\end{eqnarray*}
Furthermore, when the following condition 
\begin{eqnarray}
(p^{R_3+R_1-R}-p^{R_1})(D-1)d'-(1-p^{R_1-R})>0\label{eqn:cod}
\end{eqnarray} is also satisfied, we have 
\begin{eqnarray*}
\frac{1-p^R}{1-p}(n_1+n_2+n_3)>\sum^3_{i=1}\frac{1-p_i}{1-p}n_i
\end{eqnarray*}
\end{lem}
The proofs of Lemma \ref{lem:pckt} and \ref{lem:comp1} can be found in Appendix \ref{appen:A} and \ref{appen:B}, respectively. 

In the above lemmas, we assume that $\tilde{p}\geq p$, and we want to show that the assumption is sound.
Consider two scenarios, where the unmodified IEEE 802.11 standard and our proposed method are used, respectively. We assume that at the beginning of the video transmission, the collision probabilities are the same, i.e., $\tilde{p}(0)=p(0)$. Let $\tilde{p}(s)$ and $p(s)$ be the collision probabilities measured at the end of time interval $s$ for the two scenarios, respectively, where a time interval is the duration between two consecutive packet loss probability updates (see Line \ref{lin:12} of Algorithm \ref{alg1}) and should not be confused for the time slot defined in the IEEE 802.11 standard. Similarly, we define $M(s)$ and $M_i(s)$ which are the numbers of packets in the two scenarios, respectively, measured at the end of time interval $s$, where $i=1,2,3$. Since $M(s)$ and $M_i(s)$ are all random variables and difficult to be modeled, we take their expected values $n(s)$ and $n_i(s)$, for $i=1,2,3$ as an approximation by the method of mean-field approximation \cite{Networks}. As long as the number of packets is large, the approximation is accurate. 


During the first time interval, the collision probability $\tilde{p}(0)=p(0)$, and then, according to Lemma \ref{lem:comp1}, we have $$\frac{1-p^R(0)}{1-\tilde{p}(0)} n(1)\geq\sum^3_{i=1}\frac{1-p^{R_i}(0)}{1-p(0)}n_i(1)$$which implies that the expected number of transmission attempts in the first scenario is not less than that in the second scenario. Thus, at the end of the first time interval, the collision probability in the first scenario is also not less than that in the second scenario, i.e., $\tilde{p}(1)\geq p(1)$. As a result, in the second time interval, the expected number of transmission attempts in the first scenario is also not less than that in the second scenario, i.e.
$$\frac{1-p^R(1)}{1-\tilde{p}(1)} n(2)\geq\sum^3_{i=1}\frac{1-p^{R_i}(1)}{1-p(1)}n_i(2).$$
The same reasoning holds for the future time intervals. Thus, in our proposed method, the expected number of transmission attempts is always not greater than that in the original IEEE 802.11 standard at all time intervals, and at the end of each time interval, the collision probabilities satisfy $\tilde{p}(s)\geq p(s)$. Note that it is also consistent with the simulation results presented in Section \ref{sec:simu}.


Now we are ready to present the main result as follows. By comparing the performance of the IEEE 802.11 standard and that of our proposed method, we prove that the expected number of frozen frames is reduced when our proposed method is used.
\begin{thm}\label{thm:main}
If $\tilde{p}\geq p$ and (\ref{eqn:cod}) are satisfied, when our proposed method is used, the expected number of frozen frames is upper bounded by 
\begin{eqnarray}
N'_f<\min\left\{N_f, \frac{N_f}{[(d+(D-1)d')(1-\frac{d'-1}{2}p_1)-d]p_0+1}\right\}
\label{eqn:result}
\end{eqnarray}
\end{thm}

The proof of Theorem \ref{thm:main} contains two parts. We want to show that $N'_f$ is upper bounded by the two elements on the right hand side of (\ref{eqn:result}). In the remainder of this section, we prove them in the following two lemmas.

\begin{lem}
If $\tilde{p}\geq p$, when our proposed method is used, the expected number of frozen frames is smaller than that when the IEEE 802.11 standard is used at the video user, i.e.,
\begin{eqnarray}
N'_f<N_f.\label{eqn:bound1}
\end{eqnarray}
\end{lem}

\begin{IEEEproof}  
Denote by $N_I$ and $N'_I$ the numbers of IDR frames when the IEEE 802.11 standard and our proposed method is used at the video user, respectively. 
As we assume that every IDR frame and non-IDR frame are encoded into $d$ and $d'$ packets, respectively, the total numbers of packets when the IEEE 802.11 standard is used is given by
\begin{eqnarray*}
n&=&dN_I+d'(N-N_I)\\
&=&d'N+\Delta d N_I.
\end{eqnarray*}
Similarly, when our proposed method is used, the total number of packets is 
\begin{eqnarray*}
n_1+n_2+n_3=d'N+\Delta d N'_I.
\end{eqnarray*}
By Lemma \ref{lem:pckt}, we know that $n>n_1+n_2+n_3$. From the above two equations, we have $N_I>N'_I$.
Notice that every frozen interval triggers the generation of an IDR frame, and except the first IDR frame, which is the first frame of the video sequence, IDR frame only appears immediately after a frozen interval. Then, we have 
\begin{eqnarray*}
N_f&=&(N_I-1)D\\
N'_f&=&(N'_I-1)D.
\end{eqnarray*}
Consequently, the number of frozen frames when our proposed method is used is smaller than that when the unmodified IEEE 802.11 standard is used, i.e.,
\begin{eqnarray*}
N'_f<N_f.
\end{eqnarray*}
\end{IEEEproof}

\begin{lem}
If $\tilde{p}\geq p$ and (\ref{eqn:cod}) are satisfied, when our proposed method is used, the expected number of frozen frames is upper bounded by 
\begin{eqnarray}
N'_f<
 \frac{N_f}{[(d+(D-1)d')(1-\frac{d'-1}{2}p_1)-d]p_0+1}\label{eqn:bound2}
\end{eqnarray}
\end{lem}



\begin{IEEEproof}
In the proof of Lemma \ref{lem:comp1}, we obtain 
\begin{eqnarray}
&&\frac{1-p^R}{1-p}(n_1+n_2+n_3)-\sum^3_{i=1}\frac{1-p_i}{1-p}n_i\nonumber\\
&\geq&\frac{p^Rn_1}{1-p}[(p^{R_3+R_1-R}-p^{R_1})(D-1)d'-(1-p^{R_1-R})]\nonumber
\end{eqnarray}
where the right hand side of the above inequality is a positive increasing function of $n_1$. As the number of packets is increasing, the lower bound of the left hand side of the above inequality is also increasing and remains positive. Notice that $n$ and $n_i$ are the expected values of random variables $M$ and $M_i$. When the left hand side of the above inequality is large enough, the compatibility condition (\ref{eqn:compa_ineqn_alg}) is always satisfied. It implies that according to Algorithm \ref{alg1}, no frame with priority 2 will be generated after the beginning of the video sequence. 

As discussed above, except the beginning of the video sequence, no frame is assigned priority 2. Hence, a frame with priority 1 is followed by another frame with priority 1, when all packets of the former are transmitted successfully. Notice that according to Algorithm \ref{alg1}, the priority doesn't change within a frame. Even if a packet of a frame with priority 1 is dropped, the remaining packets of the same frame still have the same priority and the packets of the subsequent frame are then assigned priority 3. Each frozen interval contains $D-1$ subsequent frames with priority 3, each of which is encoded into $d'$ packets. The first $(D-1)d'-1$ packets are followed by another packet with priority 3 with probability 1, and the last one is followed by a packet with priority 1, which belongs to the next IDR frame, with probability 1. This process can be modeled by the discrete-time Markov chain shown in Fig. \ref{fig:markov}.
\begin{figure}[t]
    \centering
    \includegraphics[width=0.6\textwidth]{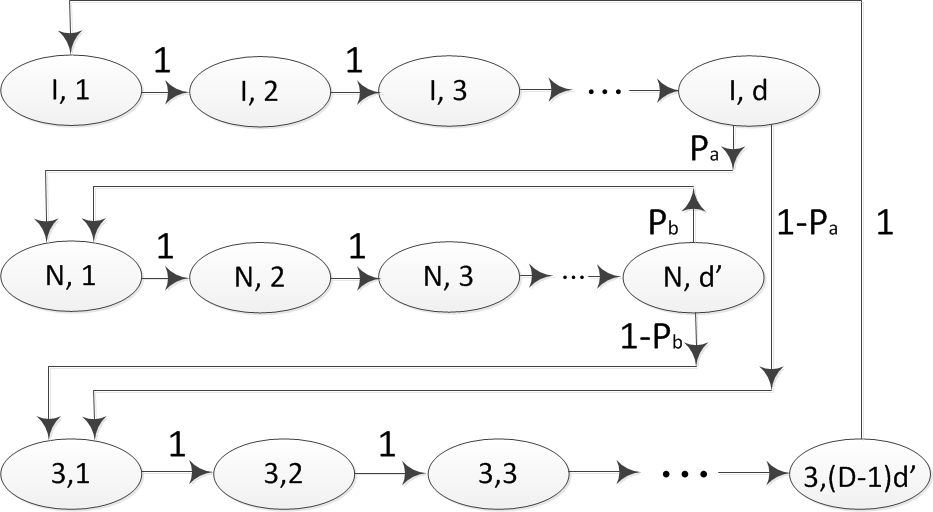}
    \caption{Markov chain model for video packet classes, where $(I,i)$, $(N,i)$ and $(3,i)$ represent the states for the $i$-th packet of an IDR frame, a non IDR frame and a frame with priority 3, respectively.}
    \label{fig:markov}
\end{figure}

In Fig. \ref{fig:markov}, the states in the last row represent the $(D-1)d'$ packets with priority 3 in each frozen interval. The states in the first two rows represent the $d$ and $d'$ packets of an IDR frame and a non-IDR frame with priority 1, respectively, where the state $(I, i)$ is for the $i$-th packet of the IDR frame, and the state $(N, j)$ is for the $j$-th packet of the non-IDR frame with priority 1. After a frozen interval, it is followed by $d$ packets of an IDR frame with priority 1. If all these $d$ packets are transmitted successfully, they are followed by $d'$ packets of a non-IDR frame, and otherwise, they initialize a new frozen interval. After the transmission of a non-IDR frame, it is followed by another non-IDR frame unless the transmission fails. Suppose that $P_a$ and $P_b$ are the probabilities that the transmissions of an IDR frame and a non-IDR frame with priority 1 are successful, respectively. The transmission of an IDR frame is successful if and only if all the $d$ packets of the IDR frame are transmitted successfully. For each packet, the packet loss rate is $p_1$, since it has priority 1. Thus, we have
\begin{eqnarray}
P_a=(1-p_1)^d.
\end{eqnarray}
For the non-IDR frames, notice that they also have priority 1. Then, the probability $P_b$ is given by
\begin{eqnarray}
P_b=(1-p_1)^{d'}.
\end{eqnarray}

When $D=1$, no frame is assigned priority 3, and then, we don't have the states in the last row in Fig. \ref{fig:markov}. If any frame is dropped in transmission, it will be followed immediately by another IDR frame. In this case, the discrete-time Markov chain becomes the model in Fig. \ref{fig:markovD1}. The following derivation is based on the model in Fig. \ref{fig:markov}. However, it is also suitable when $D=1$.
\begin{figure}[t]
    \centering
    \includegraphics[width=0.6\textwidth]{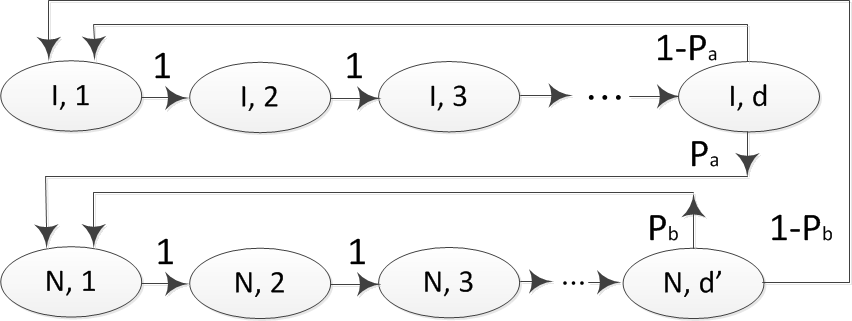}
    \caption{Markov chain model for video packet classes when $D=1$}
    \label{fig:markovD1}
\end{figure}

Let $q_{I,i}$, $q_{N,j}$ and $q_{3,k}$, for $1\leq i\leq d$, $1\leq j\leq d'$ and $1\leq k\leq (D-1)d'$, be the stationary distribution of the Markov chain. Notice that $q_{I,1}=q_{I,2}=\cdots=q_{I,d}$,  $q_{N,1}=q_{N,2}=\cdots=q_{N,d'}$ and $q_{3,1}=q_{3,2}=\cdots=q_{3,(D-1)d'}$. Furthermore, we have
\begin{eqnarray}
q_{I,1}&=&q_{3,(D-1)d'}\\
q_{N,1}&=&P_aq_{I,d}+P_bq_{N,d'}\\
q_{3,1}&=&(1-P_a)q_{I,d}+(1-P_b)q_{N,d'}
\end{eqnarray}
From the above equations, we derive that
\begin{eqnarray}
q_{N,j}&=&\frac{P_a}{1-P_b}q_{I,1}\\
q_{3,k}&=&q_{I,1}
\end{eqnarray}
Considering the normalization condition $$dq_{I,1}+d'q_{N,1}+(D-1)d'q_{3,1}=1,$$ it is not difficult to obtain that
\begin{eqnarray}
q_{I,1}=\frac{1-P_b}{[d+(D-1)d'](1-P_b)+P_ad'}.
\end{eqnarray}

Let $q_I$ be the probability that a packet belongs to an IDR frame, which is given by  $$q_I=\sum_{i=1}^{d}q_{I,i}=\frac{d(1-P_b)}{[d+(D-1)d'](1-P_b)+P_ad'}.$$
In a video sequence containing $n_1+n_2+n_3$ packets, the expected  number of packets which belong to an IDR frame is obtained by
$$n'_I=q_I(n_1+n_2+n_3).$$
From (\ref{eqn:pkt_IDR}), we have
\begin{eqnarray}\nonumber
N'_f&=&(\frac{n'_I}{d}-1)D\\\nonumber
&<&\frac{n'_ID}{d}\\\nonumber
&=&\frac{q_I(n_1+n_2+n_3)D}{d}\\\nonumber
&=&\frac{D(1-P_b)(n_1+n_2+n_3)}{[d+(D-1)d'](1-P_b)+P_ad'}\\
&<&\frac{D(1-P_b)n}{[d+(D-1)d'](1-P_b)+P_ad'}\label{eqn:result1}
\end{eqnarray}
where the last inequality follows from the fact that $n_1+n_2+n_3<n$. 

By Taylor's theorem, the probability $P_a$ can be rewritten as
\begin{eqnarray*}
P_a&=&(1-p_1)^d\\
&=&1-dp_1+\frac{d(d-1)}{2}(1-\xi)^{d-2}p_1^2
\end{eqnarray*}
where $0\leq\xi\leq p_1\leq1$. Thus, we have
\begin{eqnarray*}
1-dp_1\leq P_a\leq 1-dp_1+\frac{d(d-1)}{2}p_1^2.
\end{eqnarray*}
Similarly, we obtain that
$$d'p_1-\frac{d'(d'-1)}{2}p_1^2\leq 1-P_b\leq d'p_1.$$
Then, applying the above bounds, inequality (\ref{eqn:result1}) can be rewritten as
\begin{eqnarray}
N'_f&<&\frac{Dd'p_1n}{[d+(D-1)d'](d'p_1-\frac{d'(d'-1)}{2}p_1^2)+(1-dp_1)d'}\nonumber\\
&=&\frac{Dp_1n}{[d+(D-1)d'](p_1-\frac{d'-1}{2}p_1^2)-dp_1+1}\nonumber\\
&=&\frac{Dp_0n}{[d+(D-1)d'](p_0-\frac{d'-1}{2}p_0p_1)-dp_0+\frac{p_0}{p_1}}\nonumber\\
&=&\frac{Dp_0n}{[(d+(D-1)d')(1-\frac{d'-1}{2}p_1)-d]p_0+\frac{p_0}{p_1}}\nonumber\\
&<&\frac{N_f}{[(d+(D-1)d')(1-\frac{d'-1}{2}p_1)-d]p_0+1}\nonumber
\end{eqnarray}
where the last inequality follows from the fact that $p_0>p_1$ and $N_f=Dp_0n$.
\end{IEEEproof}

\begin{figure}[t]
    \centering
    \includegraphics[width=0.6\textwidth]{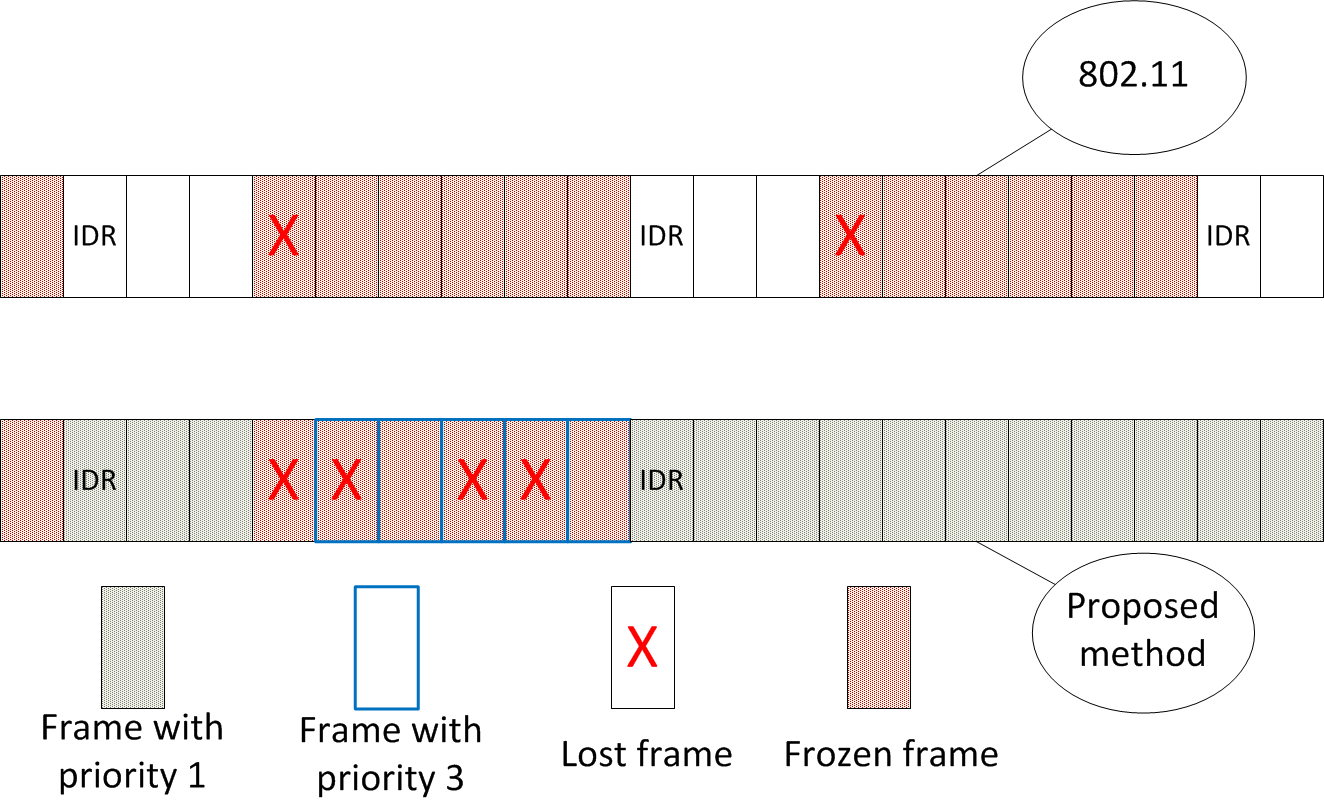}
    \caption{Frozen frame comparison between the unmodified IEEE 802.11 and the proposed method}
    \label{fig:comparison}
\end{figure}

From the result above, we have the following observations: the expected freeze time for our proposed method is always reduced compared to the unmodified IEEE 802.11 standard; the longer the frozen interval $D$, the greater the gain compared to the unmodified IEEE 802.11 standard. As shown in Fig. \ref{fig:comparison}, for the case of the unmodified IEEE 802.11 standard, after a packet loss, the video receiver shows frozen frames until the next IDR frame, regardless of whether the frames before the IDR frame are received successfully. However, for the case of our proposed method, priority 3 is assigned to the frames following a packet loss, and more packets are dropped during the error propagation. As a compensation, the frames from the next IDR frame have a higher retry limit, and as a result, a lower packet loss probability is achieved for these frames and the total number of frozen frames is reduced.  By increasing the retry limit for the high priority packet, our proposed method concentrates the packet losses into small segments of the entire video sequence to improve the video quality.

\section{Simulation Results}\label{sec:simu}

\begin{figure}[t]
    \centering
    \includegraphics[width=0.6\textwidth]{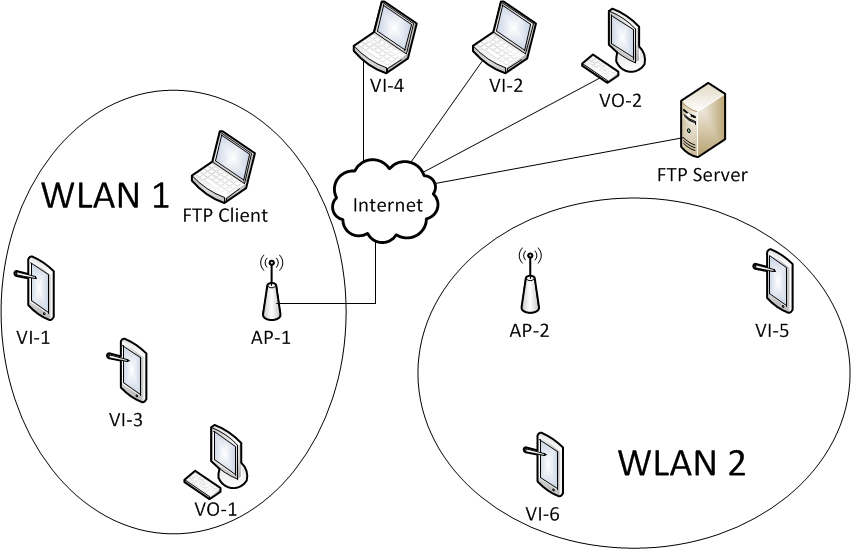}
    \caption{Network topology used in the simulations}
    \label{fig:net_topo}
\end{figure}
%

Our proposed scheme is evaluated using the network topology in Fig. \ref{fig:net_topo}, which contains a video teleconferencing session with our proposed QoE-based optimization (VI-1 and VI-2) and other cross traffic, including a voice session, an FTP session, and a video teleconferencing session without our proposed QoE-based optimization (VI-3 and VI-4). In the simulations, we only consider one-way video transmission from VI-1 to VI-2, while the video teleconferencing between VI-3 and VI-4 are two-way. VI-1 and VI-3 are in the same WLAN with the FTP Client and the voice user VO-1. The access point AP-1 communicates with VI-2, VI-4, the FTP Server, and the voice user VO-2 through the Internet, with a one-way delay of 100 ms in either direction. The H.264 video codec is implemented for VI-1 and VI-2.

Without our proposed QoE-based optimization, the retry limit $R$ for all packets is set to 7, the default value in the IEEE 802.11 standard. Three levels of video priority are assigned in video teleconferencing sessions with our proposed QoE-based optimization, and the corresponding retry limits are $(R_1, R_2, R_3)=(8, 7, 1)$, where as we discuss before, we set $R_2=R$ because it leads to a tractable analytic model and makes the implementation easier. Since larger retry limit may cause longer delay to certain packets, we only increase the retry limit $R_1$ by 1. Meanwhile, no matter whether the packets with priority 3 are received successfully or not, their corresponding frames will be frozen, so we assign the retry limit $R_3=1$ for them. At the video sender, a packet is discarded when its retry limit is exceeded. In our proposed scheme, the video receiver detects a packet loss when it receives the subsequent packets or it does not receive any packets for a time period. Then it sends the packet loss information to the video sender through RTCP, and a new IDR frame is generated after the RTCP feedback is received by the video sender. From the time of the lost frame until the next IDR frame is received, the video receiver only presents frozen frames.

Two test video sequences are transmitted from VI-1 to VI-2. One is the low motion Foreman sequence, which has a frame rate 30 frames/sec and a duration of approximately 10 seconds, containing 295 frames. The other one is the high motion Basketball Passing sequence. Its frame rate is 60 frames/sec, and the duration is 5 second, containing 300 frames. All the cross traffic is generated by OPNET 17.1. For the cross video session from VI-3 to VI-4, the frame rate is 30 frames/sec, and the outgoing and incoming stream frame sizes are both 8500 bytes.
For the TCP session between the FTP client and server, the receive buffer is set to 8760 bytes. All the numerical results in this section are averaged over 100 seeds, and for each seed, the data is collected in the durations of the video sequences.




To make the network performance comparable to those reported in ~\cite{Towsley12}, we also introduce another WLAN to increase the error probability $p$. It consists of an AP and two stations shown as WLAN 2 in Fig. \ref{fig:net_topo}. Both of these two IEEE 802.11n WLANs operate on the same channel. The data rates are 13 Mbps, and the transmit powers are 5 mW. The buffer sizes at the APs are 1 Mbits. The numbers of spatial streams are set to 1. The distances of the APs and the stations are set to enable the hidden node problem. In the simulations, the distance between the two APs is 300 meters, and the distances between VI-1 and AP-1, and between AP-2  and VI-5, are both 350 meters. A video teleconferencing session is initiated between VI-5 and VI-6 through AP-2. The frame rate is 30 frames/sec, and both the incoming and outgoing stream frame sizes are used to adjust the packet loss rate of the video teleconferencing session with our proposed method operating at VI-1.

\begin{figure}[t]
    \centering
    \includegraphics[width=0.6\textwidth]{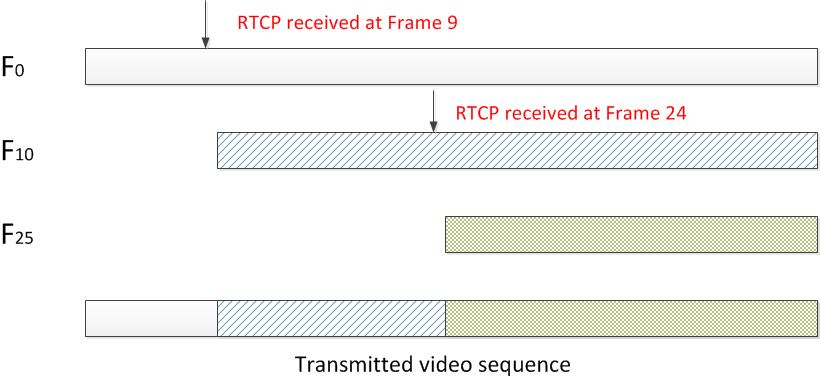}
    \caption{Real-time video transmission in OPNET}
    \label{fig:opnet}
\end{figure}

Network events affect the behavior of the video encoder, and vice versa. The former is often ignored in simulation based studies by feeding a pre-determined video sequence to the network \cite{Jiang12,Zhou13}. In our simulation study, we dynamically generate the video frames to be fed into the network according to the network events in the network. Specifically, we capture the dynamic IDR frame insertion that is triggered by reception of the packet loss feedback conveyed by RTCP packets in OPNET. The details are as follows: Let $F_n$, $n=0,1,2,\cdots$, be the video sequence beginning from frame $n$, where frame $n$ is an IDR frame and all the subsequent frames are P-frames until the end of the video sequence. We start from the transmission of video sequence $F_0$, and suppose that RTCP feedback is received when we are transmitting frame $i-1$. After the transmission of the current frame, we switch to the video sequence $F_i$, which causes the IDR frame insertion at frame $i$, and use frame $i$ and the subsequent frames of $F_i$ to feed the video sender simulated in OPNET. In Fig.\ref{fig:opnet}, we depict the video sequence, where RTCP feedback is received when frame-$9$ and frame-24 are transmitted. Note that in the OPNET simulation it is the size rather than the content of each packet that is of interest. We encode all possible video sequences $F_n$, $n=0,1,2,\cdots$, which is a one-time effort, and store the size of every packet of all video sequences. Then, in the simulations, when we receive an RTCP feedback, we switch to the appropriate video sequence.

\begin{figure}[t]
    \centering
    \includegraphics[width=0.6\textwidth]{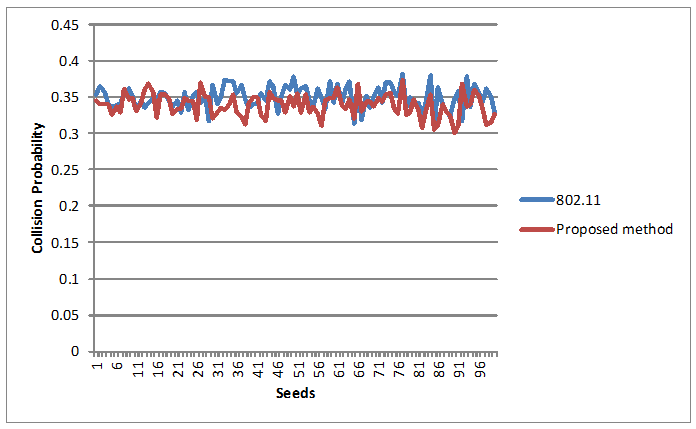}
    \caption{Collision probabilities $p$ of a single transmission attempt using the IEEE 802.11 standard and our proposed method for 100 seeds}
    \label{fig:collision_prob}
\end{figure}
In Fig. \ref{fig:collision_prob}, we present the collision probabilities $p$ for 100 seeds, when the IEEE 802.11 standard and our proposed method are used at the video user, respectively. The average collision probabilities are 0.350 and 0.339 for the IEEE 802.11 standard and our proposed method, respectively, which confirms the assumption $\tilde{p}\geq p$. 
Comparing these two values, it is reasonable to use the collision probability from our proposed method as an approximation of collision probability when only the IEEE 802.11 standard is applied.

\begin{figure}
    \centering
    \includegraphics[width=0.6\textwidth]{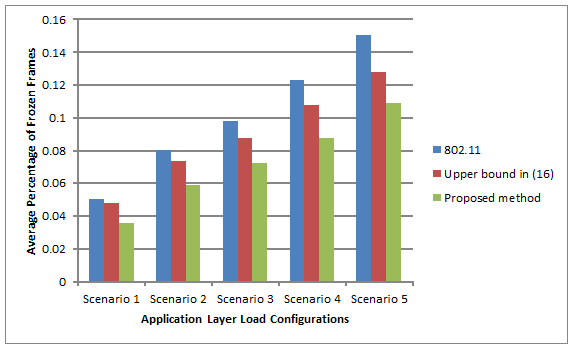}
    \caption{Average fractions of frozen frames versus different application layer load configurations for Foreman, which result in packet loss rates 0.0023, 0.0037, 0.0044, 0.0052 and 0.0058, respectively, when the IEEE 802.11 standard is used at the video user}
    \label{fig:num_frm_plr}
\end{figure}
\begin{figure}
    \centering
    \includegraphics[width=0.6\textwidth]{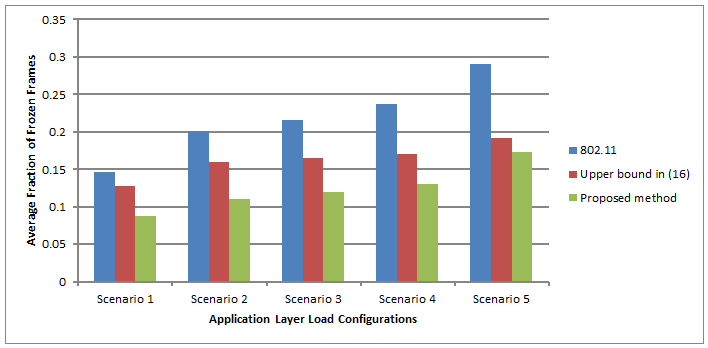}
    \caption{Average fractions of frozen frames versus different application layer load configurations for Basketball passing in the sames scenarios}
    \label{fig:num_frm_plr_bskt}
\end{figure}

In Fig. \ref{fig:num_frm_plr}, the average fractions of frozen frames using the IEEE 802.11 standard and our proposed QoE-based optimization are presented, where the Foreman sequence is transmitted. For different application layer load configurations, we tune the cross traffic between VI-5 and VI-6 to obtain different packet loss rates, when the IEEE 802.11 standard is used at the video user. The packet loss rates are  0.0023, 0.0037, 0.0044, 0.0052 and 0.0058, for scenarios 1 to 5, respectively. Then, we run the simulations using our proposed method with the same cross traffic configurations. We also show the upper bound for our proposed method in (\ref{eqn:result}), where the parameters $D$, $d$, $d'$ and $p_0$ are averaged from the simulation results, and it is confirmed that (\ref{eqn:cod}) is satisfied. In Fig. \ref{fig:num_frm_plr_bskt}, we also present the average fractions of frozen frames of Basketball Passing video for the same configurations. We observe that the average fraction of frozen frames of our proposed method is strictly less than the upper bound. As the packet loss rate increases, the average fraction of frozen frames increases regardless of whether our proposed method is used or not, and the performance of our method remains strictly better than that of the corresponding value of the IEEE 802.11 standard.

\begin{figure}[t]
    \centering
    \includegraphics[width=0.6\textwidth]{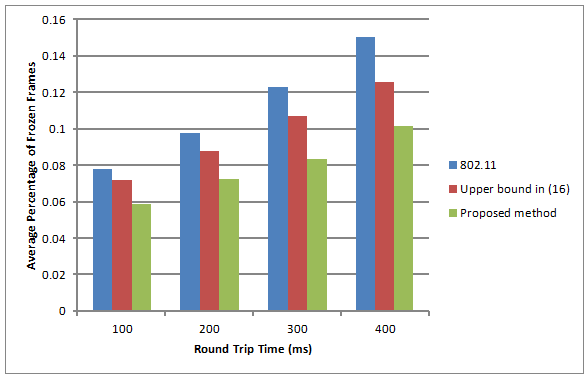}
    \caption{Average fractions of frozen frames for Foreman versus round trip times}
    \label{fig:num_frm_rtt}
\end{figure}
\begin{figure}[t]
    \centering
    \includegraphics[width=0.6\textwidth]{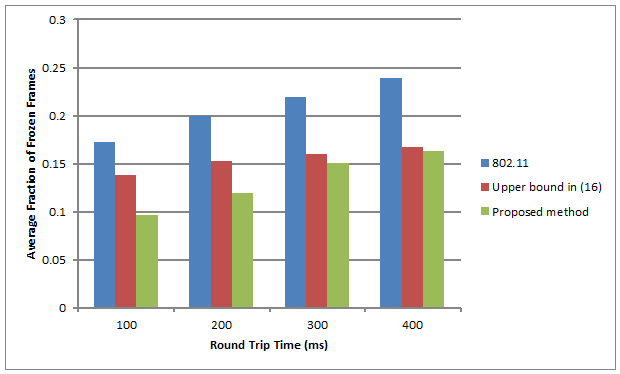}
    \caption{Average fractions of frozen frames for Basketball Passing versus round trip times}
    \label{fig:num_frm_rtt_basket}
\end{figure}
In Fig. \ref{fig:num_frm_rtt}, we show the average fractions of frozen frames for different RTTs between video sender and receiver, when the application layer load configuration 3 is applied. The Foreman sequence is transmitted. Notice that the feedback delay is at least 1 RTT between video sender and receiver. As the RTT increases, the feedback delay increases, the duration of every frozen interval increases, more frames are affected by each packet loss, and the fraction of frozen frames increases. From the upper bound in (\ref{eqn:result}), we infer that the gain of our proposed method compared to the IEEE 802.11 standard increases, when a larger RTT is applied and this is consistent with the numerical results in Fig. \ref{fig:num_frm_rtt}. When the RTT is 100 ms, the average fraction of frozen frames using our proposed method is $24.5\%$ less, compared to that using the IEEE 802.11 standard. When the RTT is 400 ms, the gain increases to $32.6\%$. Moreover, the average fractions of frozen frames using the proposed method are always less than the upper bound in (\ref{eqn:result}). Similar results are also observed for Basketball Passing video in Fig. \ref{fig:num_frm_rtt_basket}.

\begin{figure}[t]
    \centering
    \includegraphics[width=0.6\textwidth]{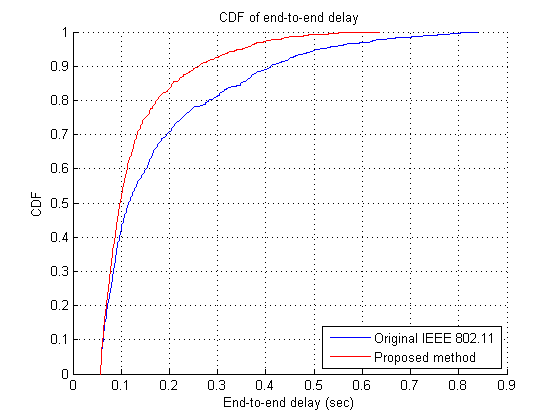}
    \caption{CDF of the packets end-to-end delays}
    \label{fig:delay}
\end{figure}

In Fig. \ref{fig:delay}, we present the CDF of the packet end-to-end delay between the video sender and receiver, when the application layer load configuration 3 is applied and the RTT is 100 ms. For our proposed method, because of Lemma \ref{lem:pckt}, less packets are generated than the original IEEE 802.11 standard. As a result, the end-to-end delay may be well reduced when our proposed method is used, which is corroborated by the results in Fig. \ref{fig:delay}.

\begin{table}[t]
\begin{center}\captionof{table}{Average throughputs for cross traffic with application layer load configuration 2}
    \begin{tabular}{ | c | c | c | c | c | c | }
    \hline
      & \multicolumn{5}{|c|}{Average Throughputs (Bytes/sec)} \\ \cline{2-6}
    & VI-3 & VI-4 & VO-1 & VO-2 & FTP \\ \hline
    IEEE 802.11   & 254962 & 255686 & 3570 & 3617 & 40732
 \\ \hline
    Proposed method  & 254766 & 255680 & 3492 & 3672 & 42985 \\ \hline
    \end{tabular}\label{tab:simu1}
\end{center}
\end{table}

\begin{table}[t]
\begin{center}\captionof{table}{Standard deviations of throughputs for cross traffic with application layer load configuration 2}
    \begin{tabular}{ | c | c | c | c | c | c | }
    \hline
      & \multicolumn{5}{|c|}{Standard Deviations of Throughputs (Bytes/sec)} \\ \cline{2-6}
    & VI-3 & VI-4 & VO-1 & VO-2 & FTP \\ \hline
    IEEE 802.11   & 9580 & 3749 & 2867 & 2808 & 29679
 \\ \hline
    Proposed method  & 10786 & 3840 & 2853 & 2887 & 29544 \\ \hline
    \end{tabular}\label{tab:simu2}
\end{center}
\end{table}

\begin{table}[t]
\begin{center}\captionof{table}{Average throughputs for cross traffic with application layer load configuration 5}
    \begin{tabular}{ | c | c | c | c | c | c | }
    \hline
      & \multicolumn{5}{|c|}{Average Throughputs (Bytes/sec)} \\ \cline{2-6}
    & VI-3 & VI-4 & VO-1 & VO-2 & FTP \\ \hline
    IEEE 802.11   & 253806 & 255598 & 3659 & 3939 & 4726
 \\ \hline
    Proposed method  & 254275 & 255687 & 3551 & 3682 & 4805 \\ \hline
    \end{tabular}\label{tab:simu3}
\end{center}
\end{table}

\begin{table}[t]
\begin{center}\captionof{table}{Standard deviations of throughputs for cross traffic with application layer load configuration 5}
    \begin{tabular}{ | c | c | c | c | c | c | }
    \hline
      & \multicolumn{5}{|c|}{Standard Deviations of Throughputs (Bytes/sec)} \\ \cline{2-6}
    & VI-3 & VI-4 & VO-1 & VO-2 & FTP \\ \hline
    IEEE 802.11   & 20420 & 4457 & 2866 & 2889 & 7502
 \\ \hline
    Proposed method  & 20396 & 4546 & 2767 & 2873 & 7416 \\ \hline
    \end{tabular}\label{tab:simu4}
\end{center}
\end{table}

In Tables \ref{tab:simu1} and \ref{tab:simu3}, we list the average throughputs for cross traffic in WLAN 1, when Foreman sequence is sent with the application layer load configurations 2 and 5 applied, respectively.   In addition, the standard deviations for these two scenarios are listed in Tables \ref{tab:simu2} and \ref{tab:simu4}, respectively. We can observe that the throughput results for the proposed method are almost the same compared to those for the IEEE 802.11 standard.

\section{Conclusion}\label{sec:conl}

We proposed a QoE-based MAC layer enhancement for WiFi that dynamically assigned different retry limits to video packets based on the packet loss events in the network subject to a compatibility design constraint. Effectively, our proposed method concentrated the packet losses into small segments of the video sequence. The number of frozen frames was reduced compared to the original IEEE 802.11 standard. Additionally simulation results showed that cross traffic was not negatively affected.

\section*{Acknowledgment}

The authors would like to thank Dr. Anantharaman Balasubramanian of InterDigital Labs for helping develop the joint video-OPNET simulation, and Dr. Rahul Vanam and Dr. Louis Kerofsky for helping set up the subjective experiment.

\appendices
\section{Proof of Lemma \ref{lem:pckt}}\label{appen:A}
\begin{IEEEproof} 
Using the original IEEE 802.11 standard, every lost packet will trigger a new IDR frame and the first frame of the video sequence is an IDR frame. So the expected total number of the IDR frames is $p_0n+1$. The expected total number of packets is given as
\begin{eqnarray*}
n=(p_0n+1)d+[N-(p_0n+1)]d',
\end{eqnarray*}
which yields
\begin{eqnarray}\label{eqn:numfrm_ori}
N=\frac{n-(p_0n+1)(d-d')}{d'}.
\end{eqnarray}

With our proposed method, every lost packet with priority 1 or 2 will cause the generation of a new IDR frame. Similar to the case with the original IEEE 802.11 standard, the expected total number of packets is given as
\begin{eqnarray*}
n_1+n_2+n_3=(p_1n_1+p_2n_2+1)d+[N-(p_1n_1+p_2n_2+1)]d',
\end{eqnarray*}which yields
\begin{eqnarray}\label{eqn:numfrm_prop}
N=\frac{(n_1+n_2+n_3)-(p_1n_1+p_2n_2+1)(d-d')}{d'}.
\end{eqnarray}

Let $\Delta d= d-d'$. From (\ref{eqn:numfrm_ori}) and (\ref{eqn:numfrm_prop}), we have
\begin{eqnarray}
n-(p_0n+1)\Delta d=(n_1+n_2+n_3)-(p_1n_1+p_2n_2+1)\Delta d.\label{eqn:numpkt}
\end{eqnarray}
Notice that $p_2=p^R\leq \tilde{p}^R=p_0$, and then, we have
\begin{eqnarray}\nonumber
(1-p_2\Delta d)(n-n_2)&\geq&(1-p_0\Delta d)n-(1-p_2\Delta d)n_2\\
&=&(1-p_1\Delta d)n_1+n_3\nonumber\\
&\geq&(1-p_1\Delta d)(n_1+n_3),\label{ineq1}
\end{eqnarray}
where the equality follows from (\ref{eqn:numpkt}), and the last inequality follows from the fact that $1-p_1\Delta d<1$. 
Since $p_1<p_2$, we have $1-p_2\Delta d<1-p_1\Delta d$. Then, it follows from (\ref{ineq1}) that
\begin{eqnarray}
n-n_2&\geq&\frac{1-p_1\Delta d}{1-p_2\Delta d}(n_1+n_3)\\
&>&n_1+n_3,
\end{eqnarray}where the first inequality follows from the fact that $p_2$ is generally very small and $p_2\Delta d$ is then less than 1.
From the above inequality, we have that $n>n_1+n_2+n_3$, i.e., for the same number of video frames, the expected number of packets when the unmodified IEEE 802.11 standard is used at the video user is greater than that when our proposed method is used.
\end{IEEEproof}

\section{Proof of Lemma \ref{lem:comp1}}\label{appen:B}


\begin{IEEEproof} From (\ref{eqn:numpkt}), we obtain that
\begin{eqnarray}
n-(n_1+n_2+n_3)=[p_0n-(p_1n_1+p_2n_2)]\Delta d. \label{eqn:numpkt2}
\end{eqnarray}Since the left hand side of (\ref{eqn:numpkt2}) is always greater than 0, we have $p_0n-(p_1n_1+p_2n_2)>0$.

Considering the compatibility condition (\ref{eqn:compat}), we have
\begin{eqnarray*}
&&\frac{1-p_0}{1-\tilde{p}} n-\sum^3_{i=1}\frac{1-p_i}{1-p}n_i\nonumber\\
&\geq&\frac{n-(n_1+n_2+n_3)-p_0n+(p_1n_1+p_2n_2+p_3n_3)}{1-p}\nonumber\\
&=&\frac{[p_0n-(p_1n_1+p_2n_2)](\Delta d-1)+p_3n_3}{1-p}\\
&\geq&0
\end{eqnarray*}
The first inequality follows from the fact that $\tilde{p}\geq p$. The equation is obtained by substituting (\ref{eqn:numpkt2}). The second inequality follows from the facts that  $p_0n-(p_1n_1+p_2n_2)>0$, $\Delta d\geq1$ and $n_3\geq0$, and the equality holds when $\Delta d=1$ and $n_3=0$. 

To prove (\ref{eqn:compat_alg}), we have
\begin{eqnarray}
\frac{1-p^R}{1-p}(n_1+n_2+n_3)-\sum^3_{i=1}\frac{1-p_i}{1-p}n_i
=\frac{1}{1-p}[(p^{R_3}-p^R)n_3-(p^R-p^{R_1})n_1]\label{eqn:31}
\end{eqnarray}
which follows from $p_1=p^{R_1}$, $p_3=p^{R_3}$ and $p_2=p^R$. From (\ref{eqn:errfrm_prop}) and (\ref{eqn:pkt3}), we obtain
\begin{eqnarray*}
n_3&=&(D-1)(p_1n_1+p_2n_2)d'\\
&\geq& (D-1)p^{R_1}d'n_1
\end{eqnarray*}

Substitute the above inequality to (\ref{eqn:31}), we have
\begin{eqnarray}
&&\frac{1-p^R}{1-p}(n_1+n_2+n_3)-\sum^3_{i=1}\frac{1-p_i}{1-p}n_i\nonumber\\
&\geq&\frac{1}{1-p}[(p^{R_3}-p^R)(D-1)p^{R_1}d'n_1-(p^R-p^{R_1})n_1]\nonumber\\
&=&\frac{p^Rn_1}{1-p}[(p^{R_3+R_1-R}-p^{R_1})(D-1)d'-(1-p^{R_1-R})]\nonumber\\
&>&0\nonumber
\end{eqnarray}
\end{IEEEproof}

\end{document}